# Improving the Testability of Object-oriented Software during Testing and Debugging Processes

Sujata Khatri
DDU College
University of Delhi, Delhi
India

R.S. Chhillar
DCSA, M.D.U
Rohtak, Haryana
India

V.B.Singh
Delhi College of Arts &
Commerce
University of Delhi, Delhi
India

## ABSTRACT

Testability is the probability whether tests will detect a fault, given that a fault in the program exists. How efficiently the faults will be uncovered depends upon the testability of the software. Various researchers have proposed qualitative and quantitative techniques to improve and measure the testability of software. In literature, a plethora of reliability growth models have been used to assess and measure the quantitative quality assessment of software during testing and operational phase. The knowledge about failure distribution and their complexity can improve the testability of software. Testing effort allocation can be made easy by knowing the failure distribution and complexity of faults, and this will ease the process of revealing faults from the software. As a result, the testability of the software will be improved. The parameters of the model along with the proportion of faults of different complexity to be removed from the software have been presented in the paper .We have used failure data of two object oriented software developed under open source environment namely MySQL for python and Squirrel SQL Client for estimation purpose.

## 1. INTRODUCTION

Software applications are fastest growing trend in the virtual world and the possibilities regarding the features and functions provided by a specific application is generating tremendous interest amongst a vast number of people around the globe. As the interest grows, so does the demand for application. Since development of large software products involves several activities which are need to be suitably coordinated to meet desired requirements. Hence in today's world the importance of developing quality software is no longer an advantage but a necessary factor. Software testing is one the most powerful methods to improve the software quality directly.

The rest of the paper is organized as follows: Section 1.1 of the paper deals with the theoretical approaches of testability. Section 1.2 gives a state of the art position of the work done in the area of measurement and improvement of testability of the procedural and object oriented software system. Software reliability growth models which is used for finding failure distribution and fault complexity of software is described in section 2. Section 3 describes Data Sets along with sample of bug reported data. failure . The impact of value of parameter of reliability growth models on improving testability is described in section 4. Conclusion of the paper and future research direction is given in section 5.

### 1.1 Testability of Software

The IEEE Standard Glossary of Software Engineering Terminology' defines testability as"(1) the degree to which a system or component facilitates the establishment of test criteria and the performance of tests to determine whether those criteria have been met, and (2) the degree to which a requirement is stated in terms that permit establishment of test criteria and performance of tests to determine whether those criteria have been met". Testability is one of the major factors determining the time and effort needed to test software system. It is costly to redesign a system during implementation or maintenance in order to overcome the lack of testability. There are so many methodologies of software development that it is difficult to list specific or stringent rules for creating testable software. Just like testing should occur from the very beginning of a project, project artifacts should be reviewed for testability from the beginning as well. The most common definition of Testability is ease *of performing testing [14]*. This definition has its roots in testing and is usually defined in terms of *observability* and *controllability*. Binder defines these two facets of testability succinctly [5]:"To test a component, you must be able to control its input (and internal state) and observe its output. If you cannot control the input, you cannot be sure what has caused a given output. If you cannot observe the output of a component under test, you cannot be sure how a given input has been processed." Based upon these definitions, it is intuitive how controllability and observability impact the ease of testing. Without controllability, seemingly redundant tests will produce different results. Without observability, incorrect results may appear correct as the error is contained in an output that you are unable to see [8].

### 1.2   Related Study on testability:

Testability is not an intrinsic property of a software artifact and cannot be measured directly (such as software size). Instead testability is an extrinsic property which results from interdependency of the software to be tested and the test goals, test methods used, and test resources (i.e., the test context)[15]. A lower degree of testability results in increased test effort. In extreme cases a lack of testability may hinder testing parts of the software or software requirements at all. Measuring testability is a challenging and most crucial task towards estimating testing efforts. Several approaches like model based testability measurement, program based testability and dependability testability assessment has been proposed. Also a number of metrics on testability measurement have been proposed, however these are applicable only in later stages of software development. Voas and Miller [4] proposed semantic metric, the \domain to range ratio (DRR): the ratio of the cardinality of the possible inputs to the cardinality of the possible outputs to indicate the testability of program early in software development process. The DRR gives important information about possible testability problems in the code required to implement specification and can help developers focus analysis and testing resources on the parts of the code that most need them. They devised a model that quantifies testability on the basis of sensitivity analysis. Gupta and Sinha[6] presented the testability measures in terms of observability and controllability and recapitulate the probe mechanism for building observability measures. Jin-Cherng, Pu-lin, & Shou-Chia [7] proposed a technique called Test Condition Oracle of software design for testability in order to reduce the testing effort and improve the software quality. The technique involves the testing characteristics of data and functions of the program into the source code to guard against the program faults. It promotes the software testability in programming step. Bruce and Haifeng [2] proposed a preliminary framework for the evaluation of software testability metric. They formulate a set of guidelines in object oriented design to improve software quality by increasing their testability.

Binder[5] did a novel work highlighting the need and significance of software testability in system development. He argued that a more testable system may provide increased reliability for a fixed testing budget. He proposed a fishbone model representing the key factors of testability.  Bruntink and Deursen [1] define and evaluate a set of metrics for assessing the testability of classes of a Java system. They evaluate a set of object oriented metric with respect to their capabilities to predict the effort needed for testing. Mouchawrab,  Briand, & Labiche [23] proposed a framework to assess testability of design modeled with the UML. They also proposed a set of operational hypotheses for each attribute that can explain its expected relationship with testability; but the hypotheses are not empirically validated. Jungmayr [24] proposes a measurement of testability from the point of view of the architecture of the system, by measuring the static dependencies between the components. Software testability can be improved during different phases of software development life cycle.

Estimating testability requires knowledge about population faults [13]. The bugs lying dormant in the software affect the progress of testing and hence testability of the software. Software faults can be introduced during any stage of software development life cycle. Software faults introduced during different phases are not of similar complexity and hence, efforts need to remove those faults are also of different level. The knowledge about remaining faults and their proportion of different complexity will help in allocating testing effort and resource management. As the knowledge about failure distribution and proportion of fault complexity grows, so does the testability of the software increases.

In the following section, we determine  fault complexity and their proportion yet to be removed from the software using reliability growth models.

## 2.  INTRODUCTION TO SOFTWARE RELIABILITY GROWTH Models

Software reliability engineering techniques play a central role in the planning and control of software development projects. In particular, it is important to document the time and nature of bug occurrences and their correction time throughout the design and implementation phases as well as during the formal testing phase. Software reliability engineering works by applying two functional ideas. First, it delivers the desired functionality for the product under development much more efficiently by quantitatively characterizing the expected use of the product and using this information to

1.  Precisely focus resources on the most used and/or most critical functions.
2.  Make testing realistically represent field conditions.

 The Software Reliability Growth Model (SRGM) is the tool, which can be used to evaluate the software quantitatively, develop test status, schedule status and monitor the changes in reliability performance. There have been many Software Reliability models developed in the last three decades. Most of these are based upon historical failure data collected during the testing phase.  These models have been utilized to evaluate the quality of the software and for future reliability predictions. They have further been used in many management decision-making problems that occur during the testing phase

A General Description of Software Reliability Growth Models

Let $\left[ N(t), t \geq 0 \right]$ denote a discrete counting process representing the cumulative number of failures experienced (fault removed) up to time $t$, i.e., $N(t)$, is said to be an NHPP with intensity function $\lambda(t)$, if it satisfies the following conditions:

I.  There are no failures experienced at time $t = 0$, i.e., $N(t = 0) = 0$ with probability 1.

II. The process has independent increments, i.e., the number of failures experienced in $(t, t + \Delta t]$, i.e., $N(t + \Delta t) - N(t)$, is independent of the history. Note this assumption implies the Markov property that the $N(t + \Delta t)$ of the process depends only on the present state $N(t)$ and is independent of its past state $N(x)$, for $x < t$.

III. The probability that a failure will occur during $(t, t + \Delta t]$ is $\lambda(t)\Delta t + o(\Delta t)$, i.e., $\Pr\left[ N(t + \Delta t) - N(t) = 1 \right] = \lambda(t) + o(\Delta t)$

   Note that the function $o(\Delta t)$ is defined as

   $$\lim_{\Delta t \to 0} \frac{o(\Delta t)}{\Delta t} = 0$$

   In practice, it implies that the second or higher order effects of $\Delta t$ are negligible.

IV. The probability that more than one failure will occur during $(t, t + \Delta t]$ is $o(\Delta t)$, i.e., $\Pr\left[ N(t + \Delta t) - N(t) > 1 \right] = o(\Delta t)$.

Based on the above NHPP assumptions, it can be shown that the probability that $N(t)$ is a given integer $k$ is expressed by

$$\Pr\left[ N(t) = k \right] = \frac{[m(t)]^k}{k!} \exp\{-m(t)\}, \quad k \geq 0$$

The function $m(t)$ is called the mean value function and describes the expected cumulative number of failures in $(0, t]$. Hence, $m(t)$ is a very useful descriptive measure of the failure behavior.

The function $\lambda(t)$, which called the instantaneous failure intensity, is defined as

$$\lambda(t) = \underset{\Delta t \to 0}{Lim} \frac{P\left[ N(t + \Delta t) - N(t) > 0 \right]}{\Delta t}$$

Given $\lambda(t)$, the mean value function $m(t) = E\left[ N(t) \right]$ satisfies

$$m(t) = \int_0^t \lambda(s)ds$$

Inversely, knowing $m(t)$, the failure intensity function $\lambda(t)$ can be obtained as

$$\lambda(t) = \frac{dm(t)}{dt}$$

Generally, by using different non-decreasing function $m(t)$, we get different NHPP models.

Define the number of remaining software failure at time t by $\overline{N}(t)$ and we have that

$$\overline{N}(t) = N(\infty) - N(t)$$

where $N(\infty)$ is the number of faults which can be detected by infinite time of testing.

It follows from the standard theory of NHPP that the distribution of $\overline{N}(t)$ is Poisson with parameter $\left[ m(\infty) - m(t) \right]$, that is

$$P\left[ \overline{N}(t) = k \right] = \frac{[m(\infty) - m(t)]^k}{k!} \exp\{m(\infty) - m(t)\}, \quad k \geq 0$$

*Notations*

$m(t)$ : Expected number of faults identified in (0,t]

$a, b$ : Constants, representing initial fault content and rate of fault removal per remaining faults for software.

p, q : proportion of dependent and independent faults

Here, we are describing some popular and basic software reliability growth models.

### 2.1 Goel-Okumoto model (Goel and Okumoto 1979)[25]

This model is the most well-known SRGM by assuming that an NHPP could describe a cumulative software failure process. Its mean value function is given as follows:

$$m(t) = a(1 - e^{-bt})$$

### 2.2 Delayed S-Shaped SRGM (Yamada, Ohba and Osaki 1983)[26]

It is observed that the MVF is often a characteristic S-shaped curve rather than the exponential growth curve [25]. In other words, the S-shapedness can be explained by considering test-efficiency improvement during the testing phases.

The MVF of the model can be defined as follows.

$$m(t) = a\{1 - (1 + bt)e^{-bt}\}$$

### 2.2 Inflection S-Shaped SRGM (Ohba 1984)[21]

Another S-shaped SRGM is proposed by Ohba assuming that some of the faults are not detectable before some other faults are removed. The MVF of Inflection S-shaped model can be presented as follows:

$$m(t) = a\left[\frac{1 - e^{-bt}}{1 + \frac{1-r}{r}e^{-bt}}\right]$$

If $r = 1$, the model reduces to the Goel-Okumoto model [1979]. For different values of $r$ different growth curves can be obtained and in that sense the model is flexible.

### 2.4 SRGM for an Error Removal Phenomenon (Kapur and Garg 1992)[27]

This model is based upon the following additional assumption: On a failure observation, the fault removal phenomenon also removes proportion of remaining faults, without their causing any failure.

Based on the assumption the fault removal intensity per unit time can be written as

$$\frac{d}{dt}m(t) = p[a - m(t)] + q\frac{m(t)}{a}[a - m(t)]$$

Solving above equation with the usual initial condition, the expected number of faults detected in $(0, t]$ is given as

$$m(t) = a\left[\frac{1 - e^{-(p+q)t}}{1 + \frac{q}{p}e^{-(p+q)t}}\right]$$

### 2.4 Software Reliability Growth Models for determining failure Distribution/fault complexity

It has been observed that any software system contains different types of faults and each type of fault requires different strategy and different testing effort to remove it. Ohba [21] proposed a hyper-exponential SRGM for a software system having different modules. Kapur et al.[17], introduced a flexible model called the generalized Erlang SRGM by classifying the faults in the software system as simple, hard and complex faults. It is assumed that the time delay between the failure observation and its removal represent the complexity of faults. Another model due to Kapur et al.[22], describes the implicit categorization of faults based on the time of detection of fault. However an SRGM should explicitly define the different types of faults as it is expected that any type of fault can be detected at any point of testing time. Kapur et al.[19] describes flexible software reliability growth model. using a power function of testing time for defining errors of different severity

In real practice, it is important to know that how many types of faults exist in the software at any time, so that different testing strategy and testing effort can be applied to remove those faults. Faults may be simple, hard, complex, more complex or even more severe. From literature survey it is clear that various software reliability growth models have been used to determine failure distribution as well as the complexity of the fault.

Achieving significant improvement over testability is easier by predicting remaining number of faults in the software and moreover their type of complexity from developer's point of view.

In this paper, we have taken two types of models, one which has been developed by a conventional approach and the other based on object oriented approach.

Now, we firstly describing the model as follows in [17]
Here, it has been assumed that different types of faults exist in the software from removal point of view due to the fact that different type of faults follows different growth curves.

$$m(t) = \sum_{i=1}^{n} a_i \left[1 - \exp(-b_i t)\left[\sum_{j=0}^{i-1}\frac{(b_i t)^j}{j!}\right]\right] \quad (1)$$

j is the number of stages to remove a fault and n is the type of fault.It is a fact that complex software system consists of different types of faults, and different types of fault need different treatment.

By incorporating learning phenomenon and assuming that the removal growth of type 1 fault which is simple in nature follows exponential curve. For other faults, which are more sever in nature, we incorporate logistic learning during removal phenomenon and these faults are depicted by different types of S-shaped curves. In the beginning, we assume that only three types of faults exist in software type 1, type 3 and type 3 (simple, hard and complex namely) and later, we extend our modeling to n types of fault as follows in [16].

$$m(t) = a_1\left(1 - \exp\left(-b_1 t\right)\right) + \frac{1}{\left(1 + \beta \exp\left(-bt\right)\right)} \sum_{i=2}^{n} a_i \left[1 - \exp\left(-bt\right)\left[\sum_{j=0}^{i-1} \frac{(bt)^j}{j!}\right]\right] \quad (2)$$

Here GE-n is model with n type of faults, $m_i(t)$ is mean number of type i faults removed in time t, $p_i$ is proportion of type i faults and $\beta_i$ is constant (for $i$ =2 to n type of faults) Now, we are describing another which is based on object oriented approach.

The following model assumes that the software system may fail due to three types of error, namely, erroneous communication between objects, erroneous execution of Private(loca1) variable/data, or erroneous execution of Public (global) variable/data (if they exist). The model further assumes that the time dependent behavior of the instruction execution follows either Exponential or Rayleigh curve, while the error removal phenomenon follows Non Homogeneous Poisson Process (NHPP) as follows in [20].

$$m(t) = ap_1\left(1 - \exp\left(-bE_1(t)\right)\right) +$$
$$ap_2\left(1 - \left(1 + bE_2(t)\right)\exp\left(-bE_2(t)\right)\right) +$$
$$ap_3\left(1 - \left(1 + bE_3(t) + \frac{\left(bE_3(t)\right)^2}{2}\right)\exp\left(-bE_3(t)\right)\right)$$

$$(3)$$

here $a = a\left(p_1 + p_2 + p_3\right)$ and

$E_1 = pE, E_2 = qE, and\ E_3 = rE$ , a is total number of

faults eventually present in the system. E is the total number of instructions executed due to accession of private, protected and public variables. p ,q, and r, are the proportion of instructions causes an accession to private , protected and public variable. $p_1, p_2$ and $p_3$ are proportion of faults due to accession of private, protected and public variables.

## 3. DESCRIPTION OF DATA SETS

**Data set-1:** SQuirreL SQL Client is a graphical SQL client written in Java that allow to view the structure of a JDBC compliant database, browse the data in tables, issue SQL commands etc. This software has been developed under open source environment www.sourceforge.net). We collected failure data of SQuirreL SQL Client from 10/3/2001(first bug reported) to 4/26/2010, during this period, 298 bugs were reported on bug tracking system as shown in table 1.

**Data set-2:** MySQL for Python software has been developed under open source environment. We collected failure data of MySQL for Python from 4/25/2001 (first bug reported) to 11/23/2009, during this period 144 bugs were reported on bug tracking system as shown in table 2.

We have considered only valid bugs which are fixed. In this section, we are describing a sample of bug reported of two object oriented software namely SQuirreL SQL Client and MySQL for Python software.

## 4. SOFTWARE RELIABILITY GROWTH MODEL PARAMETERS AND ITS IMPACT ON TESTABILITY

In this section, we are analyzing the effect of value of parameter estimates of software reliability growth models in equation (1) and (2). The value of parameters have been given in table 1 and table 2.Different software reliability growth models(GE-2 to GE-6 and GE-2(Logistic) to GE-6(Logistic) gives proportion of different types of faults lying dormant in the software(p1-p6 are proportion of faults of different complexity). These software faults are of different complexity and will take different amount of time and testing effort for removal. Table 1.1 and 2.1describes the amount of faults of different complexity yet to be remove from the software. Table 3 describes the parameter estimates of object oriented model described in equation 3 for given data sets. Table 3.1 describes the amount of faults of different complexity yet to be removed from the software.

**Table 1: Data set-1(parameter estimates)**

| Models | Parameter Estimates | | | | | | | | |
|---|---|---|---|---|---|---|---|---|---|
| | a | B | $p_1$ | $p_2$ | $p_3$ | $p_4$ | $p_5$ | $p_6$ | $\beta$ |
| GE-2 | 511 | .105 | .088 | .912 | - | - | - | - | - |
| GE-3 | 408 | .200 | .134 | .000 | .866 | - | - | - | - |
| GE-4 | 368 | .292 | .121 | .064 | .000 | .815 | | | |
| GE-5 | 555 | .332 | .052 | .163 | .010 | .000 | .775 | | - |
| GE-6 | ** | | | | | | | | - |
| GE-2(Logistic) | 313 | .383 | .194 | .806 | - | - | - | - | 25.23 |
| GE-3(Logistic) | 316 | .379 | .197 | .536 | .267 | - | - | - | 20.96 |
| GE-4(Logistic) | 321 | .393 | .191 | .515 | .000 | .295 | - | - | 19.47 |
| GE-5(Logistic) | 325 | .414 | .181 | .563 | .000 | .256 | - | - | 22.57 |
| GE-6(Logistic) | 325 | .414 | .181 | .563 | .000 | .256 | .000 | - | 22.57 |

**Table 1.1 (proportion of different complexity of faults yet to be remove from software)**

| Model | Total faults Detected | Faults to be removed | p1 | p2 | p3 | p4 | p5 | p6 |
|---|---|---|---|---|---|---|---|---|
| GE-2 | 511 | 213 | 18 | 196 | - | - | - | - |
| GE-3 | 408 | 110 | 14 | 0 | 95 | | - | - |
| GE-4 | 368 | 70 | 8 | 4 | 0 | 57 | - | - |
| GE-5 | 555 | 257 | 13 | 41 | 2 | 0 | 199 | |
| GE-6 | ** | | | | | | | |
| GE-2(Logistic) | 313 | 15 | 3 | 12 | - | - | - | - |
| GE-3(Logistic) | 316 | 18 | 3 | 10 | 5 | - | - | - |
| GE-4(Logistic) | 321 | 23 | 4 | 11 | 0 | 6 | - | - |
| GE-5(Logistic) | 325 | 27 | 4 | 15 | 0 | 0 | 6 | - |
| GE-6(Logistic) | 325 | 27 | 4 | 15 | 0 | 0 | 6 | 0 |

** Model does not give parameter estimates

**Figure 1:**

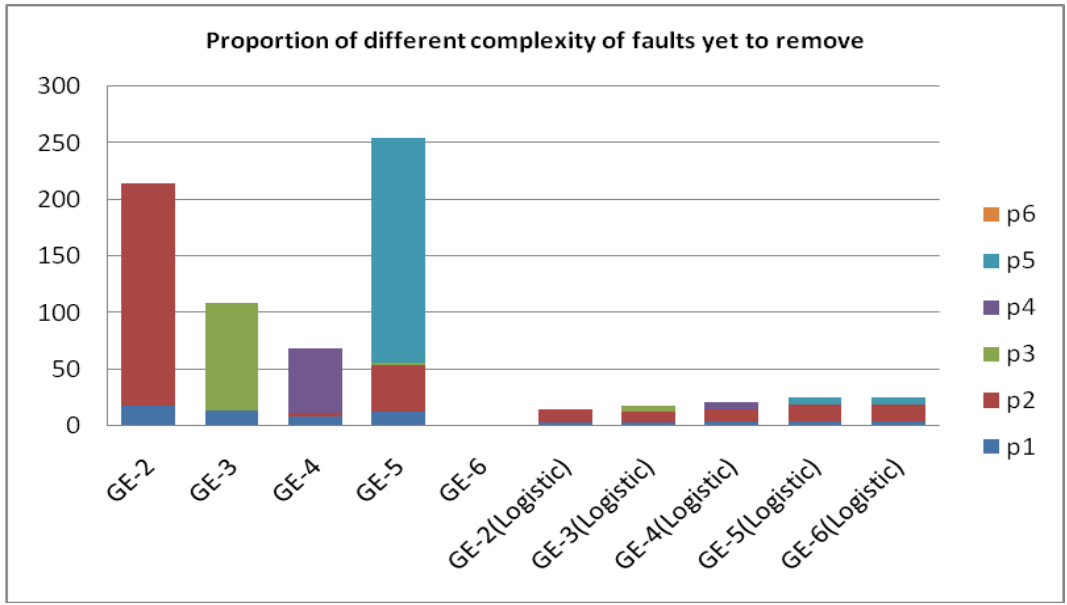

**Table 2: Data set 2(parameter estimates)**

| Models | Parameter Estimates | | | | | | | | |
|---|---|---|---|---|---|---|---|---|---|
| | a | B | $p_1$ | $p_2$ | $p_3$ | $p_4$ | $p_5$ | $p_6$ | $\beta$ |
| GE-2 | 489 | .110 | .17 | .419 | .581 | - | - | - | - |
| GE-3 | 208 | .251 | .23 | .000 | .768 | - | - | - | - |
| GE-4 | 351 | .328 | .16 | .041 | .019 | .784 | - | - | - |
| GE-5 | 171 | .513 | .19 | .022 | .088 | .000 | .701 | | - |
| GE-6 | 404 | .305 | .14 | .079 | .023 | .759 | .000 | .000 | - |
| GE-2(Logistic) | 164 | .376 | .24 | .758 | - | - | - | - | 15.8 |
| GE-3(Logistic) | 164 | .376 | .242 | .758 | .000 | - | - | - | 15.8 |
| GE-4(Logistic) | 164 | .376 | .24 | .758 | .000 | .000 | - | - | 15.8 |
| GE-5(Logistic) | 164 | .376 | .24 | .758 | .000 | .000 | .000 | - | 15.8 |
| GE-6(Logistic) | 164 | .376 | .24 | .758 | .000 | .000 | .000 | .000 | 15.8 |

**Table 2.1(proportion of different complexity of faults yet to be remove from software)**

| Models | Total faults Detected | Faults to be removed | p1 | p2 | p3 | p4 | p5 | p6 |
|---|---|---|---|---|---|---|---|---|
| GE-2 | 489 | 345 | 58 | 143 | - | - | - | - |
| GE-3 | 208 | 64 | 15 | 0 | 49 | - | - | - |
| GE-4 | 359 | 207 | 32 | 9 | 4 | 162 | - | - |
| GE-5 | 171 | 27 | 5 | 0 | 2 | 0 | 19 | - |
| GE-6 | 404 | 260 | 36 | 20 | 5 | 197 | 0 | 0 |
| GE-2(Logistic) | 164 | 20 | 49 | 15 | - | - | - | - |
| GE-3(Logistic) | 164 | 20 | 4 | 15 | 0 | - | - | - |
| GE-4(Logistic) | 164 | 20 | 4 | 15 | 0 | 0 | | - |
| GE-5(Logistic) | 164 | 20 | 4 | 15 | 0 | 0 | 0 | - |
| GE-6(Logistic) | 164 | 20 | 4 | 15 | 0 | 0 | 0 | 0 |

**Figure 2:**

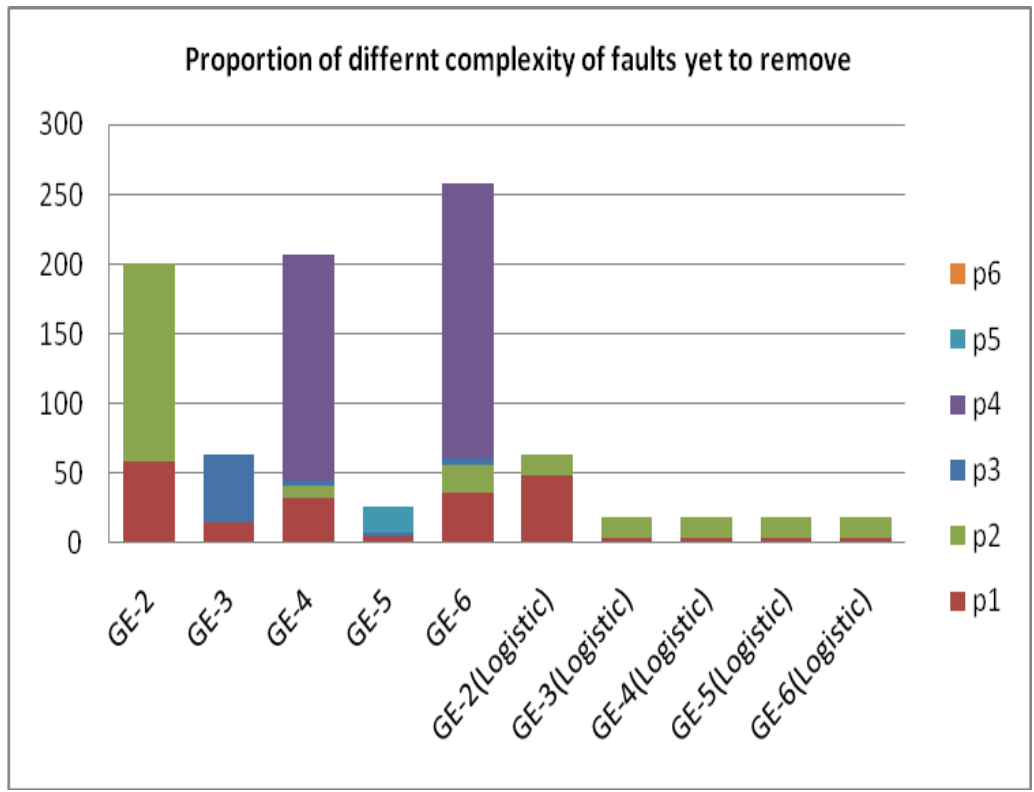

**Table 3 for Data set-1 and Data set-2(parameter estimates)**

| Parameter Estimates of (Equation-3) | Data Set-1 | | Data Set-2 | |
|---|---|---|---|---|
| | Exponential | Rayleigh | Exponential | Rayleigh |
| a | 474 | 306 | 200 | 161.905 |
| b | .017 | .091 | .036 | .090 |
| $p_1$ | .072 | .196 | .088 | .232 |
| $p_2$ | .00 | .216 | .000 | .275 |
| $p_3$ | .928 | .588 | .912 | .493 |
| p | .742 | .751 | .730 | .713 |
| q | .000 | .146 | .116 | .181 |
| R | .258 | .103 | .154 | .106 |

**Table 3.1** (Proportion of different complexity of faults yet to be removed from software)

| Proportion of faults yet to remove from software | Data Set-1 | | Data Set-2 | |
|---|---|---|---|---|
| | Exponential | Rayleigh | Exponential | Rayleigh |
| Faults yet to remove from software | 176 | 8 | 56 | 17 |
| $p_1$ | 13 | 2 | 5 | 4 |
| $p_2$ | 0 | 2 | 0 | 5 |
| $p_3$ | 163 | 5 | 51 | 8 |

Figure 3 depicts graphical presentation of different complexity of faults yet to be removed from the software.

**Figure 3:**

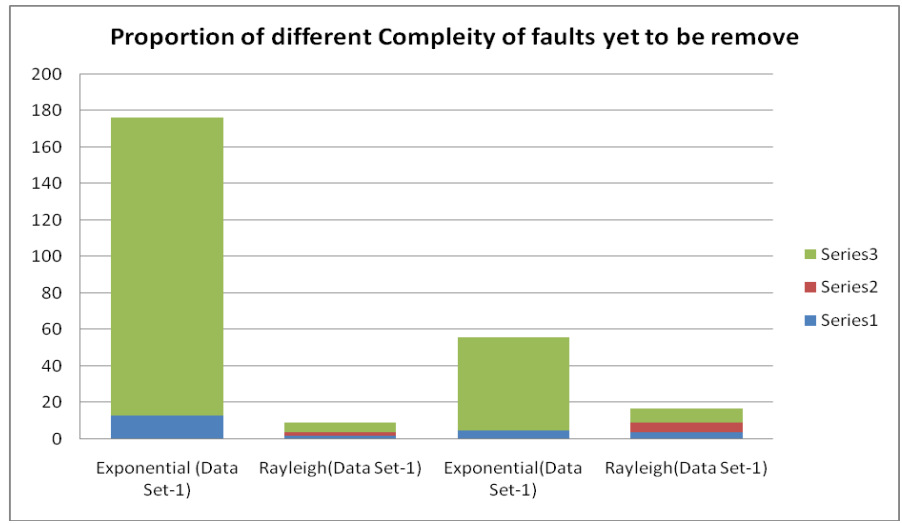

Table 1: Sample of bug reported data of SQuirreL SQL Client

| ID | Summary | Status | Opened | Assignee | Submitter | Resolution | Priority |
|---|---|---|---|---|---|---|---|
| 418713 | Python 1.5.2 adds an L | Closed | 4/25/2001 | Nobody | nobody | Wont Fix | 5 |
| 419004 | _mysql_timestamp_converter | Closed | 4/26/2001 | Adustman | nobody | Fixed | 5 |
| 424878 | Date_or_None | Closed | 5/17/2001 | Adustman | nobody | Fixed | 5 |
| 440332 | Need to #ifdef around things | Closed | 7/11/2001 | Adustman | ads | Fixed | 5 |
| 440327 | Setup.py configuration for my platform | Closed | 7/11/2001 | Adustman | gimbo | Wont Fix | 5 |
| 442299 | core-dump. Python2.1,config_pymalloc | Closed | 7/18/2001 | Adustman | nobody | Fixed | 5 |
| 45489 | Exceptions don't follow DB-API v2.0 | Closed | 7/28/2001 | Adustman | nobody | Fixed | 5 |
| 464875 | Limit bug in ZMySQLDA | Closed | 9/25/2001 | Adustman | nobody | Wont Fix | 5 |
| 464873 | Limit bug | Closed | 9/25/2001 | Nobody | nobody | Wont Fix | 5 |

Table 2: Sample of bug reported data of MySQL for Python Software

| ID | Summary | Status | Opened | Assignee | Submitter | Resolution | Priority |
|---|---|---|---|---|---|---|---|
| 467386 | Remember state of tree when refreshing | Closed | 10/3/2001 | gmackness | Colbell | Fixed | 5 |
| 467979 | Cannot type sql undr JDK1.4 | Closed | 10/4/2001 | Colbell | Nobody | Fixed | 5 |
| 472539 | wrong path separator | Closed | 10/18/2001 | Colbell | Nobody | Fixed | 5 |
| 474592 | SQL Exception with Sybase ASE 12 | Closed | 10/24/2001 | Colbell | diegodalcero | Fixed | 5 |
| 489985 | Cannot 'Cancel' from a SQL query | Closed | 12/6/2001 | Colbell | gmackness | Fixed | 5 |
| 520500 | Databased not displayed in object browsr | Closed | 2/20/2002 | Colbell | Nobody | Fixed | 5 |
| 525621 | SQL History | Closed | 3/4/2002 | Colbell | Dmishee | Fixed | 5 |
| 526656 | Incorrect linefeed replacement | Closed | 3/6/2002 | Colbell | Tlarsen | Fixed | 5 |
| 526989 | skinLF and kunstoff not available | Closed | 3/7/2002 | Colbell | nobody | Fixed | 5 |

## DATA ANALYSIS

**Table 1** shows the estimated parameter results of the models for Data Set – 1. GE-2 estimated the presence of two types of faults, GE-3, GE-4 and GE-5 estimates the presence of three types of faults and majority of them are nth type. In case of logistic removal rate GE-2 estimated the presence of two types of faults and GE-3, GE-4, GE-5 and GE-6 estimated the presence of three types of faults, however GE-6(without logistic removal rate) is not working on this data set.

**Table 1.1** presents number of faults yet to be removed from software. In model GE-2 18 faults are of type 1 and 196 faults are of type 2 which is to be removed, Model GE-3 shows 14 faults are of type 1 and 96 faults are of type 3 to be removed . GE-4 shows 8 faults of type1, 4 faults of type 2 and 57 faults are of type 4 which is to be removed. GE-5 shows 13 faults of type1, 41 faults of type 2, 2 faults of type 3 and 199 faults are of type 5 which is to be removed. GE-6 does not give parameter estimates for data set-1.

 In case of logistic removal rate GE-2 shows 3 faults are of type 1 and 12 are of type 2 to be removed from Data set-1. GE-3 presents 3 faults are of type1, 10 faults are of type 2, and 5 are of type 3 to be removed. GE-4 shows 4 faults of type 1, 11 faults of type 2, and 6 faults of type 3 to be removed. GE-5 shows 4 faults are of type 1, 15 faults of type 2 and 6 faults are of type 5 to be removed. Model GE-6 gives same result as GE-5. Graphical representation of this table is shown in figure 1.

**Table 2** shows the estimated parameter results of the models for Data Set-2. GE-2, GE-3 estimated the presence of two types of faults, GE-4, GE-5 and GE-6 estimate the presence of three types of faults and majority of them are nth type. In case of logistic removal rate GE-2, GE-3, GE-4, GE-5 and GE-6 esti estimated the presence of two types of faults and majority of t of them are nth type.

**Table 2.1** shows number of faults to be removed from Data set-2. Model GE-2 shows 58 faults are of type 1 and 143 are of type 2 to be removed from the software. GE-3 shows 15 faults of type1 and 49 faults are of type 3 which is to be removed from Data Set – 2. GE-4 shows 32 faults of type1, 9 faults of type 2, 4 faults of type 3 and 162 faults are of type 4 which is to be removed. GE-5 shows 5 faults of type 1, 2 faults of type 3 and 19 faults are of type 5 to be removed from Data Set – 2 .GE-6 shows 36 faults of type 1, 20 faults of type 2 , 5 faults of type 3 and 197 faults are of type 4 to be removed from Data set –2. Graphical representation of this table is shown in figure 2.

**Table 3** gives parameter estimates of equation (3). This table depicts the potential number of faults lying in the software and their proportion. It also gives different proportion of faults of various complexity generated due to accession of private, protected and public variables for exponential and Rayleigh type of instruction execution. Table also gives proportion of instructions executed for private, protected and public variables.

**Table 3.1** shows proportion of faults to be removed from the software. Table shows 176 faults are yet to be removed from software due to exponential type of instruction execution and 8 faults due to Rayleigh type of instruction execution for data set-1. For exponential 13 faults of type 1 and 163 faults of type 3 yet to be remove from software. For Rayleigh 2 faults of type 1, 2 faults of type 2 and 5 faults of type 3 yet to be remove from the software. For data Set-2, 56 faults are yet to be removed from software due to exponential type of instruction execution and 17 faults due to Rayleigh type of instruction execution for data set-2. For Exponential 5 faults of type 1 and 51 faults of type 3 yet to be remove. For Rayleigh 4 faults of type 1, 5 faults of type 2 and 8 faults of type 3 yet to be removed from the software.

This information about software during testing and debugging process will help project manager in deciding the allocation of effort expenditure. Based on this information, the project manager can revise his testing and debugging strategies. It will ease the process of revealing faults from the software. Therefore, testability of software can be improved.

## 5. CONCLUSION

 We have presented a novel approach for improving testability of the software using software reliability growth models. We have discussed different theoretical approach for improving and measuring testability of software. The value of parameter estimates of models for given data set has been presented in the paper. The impact and applications of parameter estimates has been shown in improving the testability of software. The knowledge of proportion of bug complexity helps in improving testability. It will help the project manager in allocating testing efforts and testing tools. The proposed testability measure can be result in higher fault detection and can also be used for the determination of modules that are more vulnerable to hidden faults. In this paper we haven't given the quantitative measure of improvement of testability but it has been shown that prior knowledge of proportion of fault of different complexity lying dormant in the software can ease the process of revealing faults. As the knowledge about failure distribution and proportion of fault complexity grows, so does the testability of the software increases.

 *Future Research Direction:*

(i)       How to quantify the improvement of testability using reliability growth models.
(ii)      How the failure distribution and fault complexity can be linked to software architecture.
We will take care of these aspects in future research papers.

## 6. ACKNOWLEDGEMENT

Corresponding author acknowledges with thanks the financial support provided by University Grants Commission, India under the project No.F.8-1(77)2010(MRP/NRCB).